\documentclass{llncs}
\usepackage{llncsdoc}
\usepackage{graphicx}
\usepackage{multirow}
\usepackage{multicol}
\usepackage{booktabs}

\usepackage[english]{babel}

\usepackage{color}

\begin{document}
\newcounter{save}\setcounter{save}{\value{section}}
{\def\addtocontents#1#2{}%
\def\addcontentsline#1#2#3{}%
\def\markboth#1#2{}%
\title{Dissimilar Symmetric Word Pairs in the Human Genome}
\titlerunning{<Dissimilar Symmetric Word Pairs>}
%
\author{Ana Helena Tavares\inst{1}
\and Jakob Raymaekers\inst{2}
\and Peter J. Rousseeuw\inst{2}
\and \\Raquel M. Silva \inst{3,4}
\and Carlos A. C. Bastos\inst{4,5}
\and Armando Pinho\inst{4,5}
\and Paula Brito\inst{6}
\and Vera Afreixo\inst{1,3,4}}
\authorrunning{Tavares et al.} 
\institute{
Department of Mathematics \& CIDMA, University of Aveiro, Portugal
\and
Department of Mathematics, KU Leuven, Belgium
\and
Department of Medical Sciences \& iBiMED, University of Aveiro, Portugal
\and
IEETA, University of Aveiro, Portugal
\and
DETI, University of Aveiro, Portugal
\and
FEP \& LIAAD - INESC TEC, University of Porto, Portugal
}
\maketitle
\begin{abstract}
In this work we explore the dissimilarity between symmetric word pairs, by comparing the inter-word distance distribution of a word to that of its reversed
complement. We propose a new measure of dissimilarity between such distributions. Since
symmetric pairs with different patterns could point to evolutionary features, we
search for the pairs with the most dissimilar behaviour. We focus our study on the
complete human genome and its repeat-masked version.
\end{abstract}

%
\section{Introduction}
Chargaff's second parity rule states that within a single strand of DNA the number of
complementary nucleotides is similar \cite{forsdyke2000}. %
An extension of this rule says that the frequency of an oligonucleotide  should be
similar to that of its reversed complement (the word obtained by reversing its letters and
interchanging $A$--$T$ and $C$--$G$). This phenomenon is known as single strand
symmetry.
Several authors discuss the prevalence of Chargaff's second parity rule (e.g.,
\cite{afreixo2013breakdown,afreixo2015analysis,albrecht2006,baisnee2002}). Various lines of
research are being explored in an attempt to explain its cause.
One approach postulates that the phenomenon would be an original feature of the
primordial genome, the most primitive nucleic acid genome, and the maintenance of strand
symmetry would rely on evolution mechanisms \cite{zhang2010strand}.

The similarity between the number of occurrences of symmetric word pairs in one strand
of DNA can be verified using frequency analysis. However, two words with the same
frequency in a sequence may exhibit very distinct distributions along that sequence.
This leads to the natural question how both words are distributed along the DNA
sequence. Are their distributions similar?

If we constrain a random generator of sequences to respect single strand symmetry
(e.g., using a high-order Markov process), it is expected that the distance distribution
of a word be similar to that of its reversed complement. %
A reasonable hypothesis is that the distance distribution of symmetric
pairs should usually be similar, and that strong deviations may have a biological origin. %

As the word length increases, more unexpected patterns may be observed in the inter-word
distance distributions, which may result in increased dissimilarity between
symmetric pairs. The similarity between distance distributions of symmetric word pairs
of length $k\leq 5$ was studied in \cite{Tavares2015}. %
For such short words the dissimilarity between symmetric pairs was
basically negligible.

This work focuses on the dissimilarity between distance distributions of symmetric pairs
of length $k=7$ in the human genome. %
We propose a new dissimilarity measure between such distributions, based on the gap
between the locations of their peaks and the difference between the sizes of these
peaks. %

The paper is organized as follows. In Section 2 we introduce a new dissimilarity measure
between distributions based on their peaks. Section 3 then identifies and investigates
the symmetric word pairs with the most dissimilar distance distributions, and Section 4
concludes.

%
\section{Methods}
\subsection{Materials}

In this study, we used the complete genome assembly (GRCh38.p2) downloaded from the
website of the National Center for Biotechnology Information. We also investigate how
well our results hold up in a masked sequence,
which excludes major known classes of repeats~\cite{lander2001}.
We used pre-masked data, available from UCSG Genome Browser
(http://genome.ucsc.edu/index.html), in witch the repeats determined by Repeat
Masker~\cite{repeatMasker} and Tandem Repeats Finder~\cite{benson1999} are replaced by N's.
The chromosomes were processed as separate sequences and non-ACGT symbols were
used as sequence separators.
Distance distributions of words were generated
using the C language. To compute the dissimilarity measures and perform the statistical
analysis, the R language was used.

\subsection{Inter-word distance distribution} 

A genomic word (or oligonucleotide) $w$ is a subsequence
in the nucleotide
alphabet $\mathcal{A}=\{A,C,G,T\}$. Words of length $k$ are elements of $\mathcal{A}^k$.
The inter-word distance sequences are defined as the
lags between the positions of the first symbol of consecutive occurrences of that
word. For instance, in the DNA segment
$A\underline{CG}T\underline{CG}ATC\underline{CG}TG\underline{CG}\,\underline{CG}$, the
inter-$CG$ distance sequence is $(3,5,4,2)$.

The inter-$w$ distance distribution (or simply distance distribution of $w$)
gives the relative frequency of each inter-$w$ distance and is denoted by $f^w$.

\subsection{Dissimilarity Measure} 

The distance distributions may present several peaks, i.e., distances with frequencies
much higher than the global tendency of the distribution. In general, the strongest
peaks occur at short distances, whereas peaks at longer distances have lower frequencies.
Looking only for the highest frequencies would not capture such local maxima.
In what follows we will take that effect into account.

\vspace{-0.35cm}
\subsubsection{Identifying peaks.} 
To determine peaks we slide a window of fixed width $h$ along the domain of the distribution.
In each such interval of width $h$ we average the absolute values of the
differences between successive frequencies, and call the result the
(average) {\it size} of the peak on that interval.
The peak's {\it location} is defined as the midpoint of the interval.
The strongest peak is then determined by the interval with the highest
size.
For the second strongest peak we only consider intervals that do not
overlap with the first one, and so on.

\vspace{-0.35cm}
\subsubsection{Dissimilarity between peaks.} 
To measure the dissimilarity between two peaks $p_1$ and $p_2$ of the same
distribution we consider the difference between their sizes and
between their locations.
We will use the following measure:
\begin{equation}
d_1(p_1,p_2)=\left(\frac{|l_1-l_2|}{R}+1\right)\left(\frac{|v_1-v_2|}{v}+1\right)-1
\label{eq:distC}
\end{equation}
where $l_1$ denotes the location and $v_1$ denotes the size of peak $p_1$
(and similarly for $p_2$).
Note that we standardize $|l_1-l_2|$ by the range $R$ of the domain
of the distribution, and $|v_1-v_2|$ by the size $v$ of its strongest
peak.
In general, the dissimilarity given by equation~(\ref{eq:distC}) increases with both the location difference and the size difference.
If the peaks have the same location the dissimilarity is reduced to a relative size
difference $|v_1-v_2|/v \leq 1$, and if they have the same size it is reduced
to a relative
location difference $|l_1-l_2|/R \leq 1$. %
When $p_1=p_2$ equation~(\ref{eq:distC}) becomes 0, and in general it
takes values between 0 and 3.

Now consider two different words $w$ and $\bar{w}$ and let $f^w$
and $f^{\bar{w}}$ be their distance distributions, defined on the same domain with length $R$. Let $p^w_i=(l_i,v_i)$ and
$p^{\bar{w}}_j=(\bar{l_j},\bar{v_j})$ be peaks in each.
To measure the dissimilarity between these peaks we propose to use
\begin{equation}
d_2(p^w_i, p^{\bar{w}}_j)=\left(\frac{|l_i-\bar{l_j}|}{R}+1\right)\left(\frac{|v_i-\bar{v_j}|}{\min\{v,\bar{v}\}}+1\right)-1
\label{eq:distD}
\end{equation}
where $v$  and $\bar{v}$ are the highest peak sizes observed in each distribution.
The denominator $\min\{v,\bar{v}\}$ yields a high dissimilarity
when one distribution has strong peaks and the other doesn't.

Note that~(\ref{eq:distD}) satisfies the semimetric property:
it reduces to zero
when the two peaks have the same location and size, and is symmetric and
non-negative. This makes it quite effective.
When $f^w =f^{\bar{w}}$
it reduces to equation~(\ref{eq:distC}).

\vspace{-0.35cm}
\subsubsection{Dissimilarity between distributions.} 
To measure the dissimilarity between two distributions we
compare their $n$ strongest peaks, for fixed $n$. We propose
\begin{equation}
d(f^w, f^{\bar{w}})=\min_{\pi\in\mathcal{P}_n}
     \{\,\sum_{i=1}^n d_2(p^w_i, p^{\bar{w}}_{\pi(i)}) \,\}
\label{eq:d}
\end{equation}
where $\mathcal{P}_n$ denotes the set of permutations of $n$ elements.
The proposed
measure~(\ref{eq:d}) depends on $n$, the number of peaks considered,
and on the
bandwidth $h$ used in the peak search.
Note that~(\ref{eq:d}) is a semimetric too.
%

\section{Results and Discussion}

There are $4^7$=16384 distinct genomic words of length $k=7$,
corresponding to 8192 symmetric word pairs.
We restrict our distance distributions 
from $k+1$ to 1000
(some distances from 1 to $k$ may be absent due to the word structure).
The dissimilarity measure~(\ref{eq:d}) between distance distributions
is computed
with bandwidth $h=5$ and the $n=3$ strongest peaks (for $n=4,\ldots,7$
we obtained similar results in much higher computation time).

Some words $w$ of length $k=7$ have a distance distribution with low
total absolute frequency $S^{w}$, so in our analysis we exclude
symmetric pairs
in which at least one word has $S^w$ below the first quartile of
$S=\{S^w: w\in \mathcal{A}^k\}$.

\subsection{Complete Genome Assembly} 
In the complete genome this first quartile is 1498, so we exclude
the symmetric pairs with $\min\{S^w, S^{\bar{w}}\} \leq 1498$
(see table~\ref{tab:stats}) and measure the dissimilarity~(\ref{eq:d})
in the remaining 6054 symmetric pairs. Let $D$ be the set formed by
these 6054 dissimilarity values.

We then automatically select the symmetric pairs with dissimilarity
under 0.129, the $10^{th}$ percentile of $D$, and those above 12.638,
its $90^{th}$ percentile.

\begin{table}[ht]
\vspace{-0.15cm}
  \centering \caption{Sum of distance frequencies $S^w$, their maximal
	ratio $\max \{ S^w/S^{\bar{w}} \}$ in a symmetric pair,
  and dissimilarity $d(f^w,f^{\bar{w}})$ inside a symmetric pair,
	for the complete genome and the masked genome.
	Results are given for all 8192 symmetric pairs and for those
	with $\min \{ S^w, S^{\bar{w}} \}$ above its first quartile.}
\renewcommand{\arraystretch}{1.2}
\scriptsize
    \begin{tabular}{lrrrrrrrrrrr}
    \toprule
           & \multicolumn{5}{c}{complete sequence}      &        & \multicolumn{5}{c}{masked sequence} \\
    \cline{2-6}\cline{8-12}
           & \multicolumn{2}{c}{all  pairs} & \multicolumn{1}{c}{} & \multicolumn{2}{c}{6054 pairs} &        & \multicolumn{2}{c}{all  pairs} & \multicolumn{1}{c}{} & \multicolumn{2}{c}{6075 pairs} \\ \cline{2-3}\cline{5-6}\cline{8-9} \cline{11-12}
           & $S^w$     & $\max\{\frac{S^w}{S^{\bar{w}}} \}$ &      & $d(f^w,f^{\bar{w}})$    & $\max\{\frac{S^w}{S^{\bar{w}}} \}$ &      & $S^w$  & $\max\{\frac{S^w}{S^{\bar{w}}} \}$ &      & $d(f^w,f^{\bar{w}})$    & $\max\{\frac{S^{w}}{S^{\bar{w}}} \}$ \\ \cline{2-3}\cline{5-6}\cline{8-9} \cline{11-12}
    Min    &  10	 & 1.000  & $\:$ & 0.003  & 1.000  & $\:\:$ & 3   & 1.000   & $\:$ &  0.032	& 1.000 \\
    Q1     &  1498	 & 1.012  &      & 0.350  & 1.009  &      & 546 & 1.015   &      & 0.507	& 	1.011 \\
    Med    &  11850	 & 1.037  &      & 0.915  & 1.022  &      & 2771 & 1.039  &      & 0.832	& 	1.026 \\
    Q3     &  28510	 & 1.165  &      & 2.936  & 1.075  &      & 6265 & 1.112  &      & 1.471	& 	1.055 \\
    Max    & 927376  & 86.74   &      & 178.7   & 83.29  &      & 277460 & 14.64&      & 21.19  & 	2.041 \\
    \bottomrule
    \end{tabular}%
  \label{tab:stats}%
\vspace{-0.1cm}
\end{table}%

The symmetric pairs with low values of~(\ref{eq:d}) have very similar
distributions.
For some words this dissimilarity is surprisingly low in spite
of their distributions having some strong peaks, which are almost
the same in the distribution of their reversed complement,
as illustrated in Figure~\ref{fig:complete}(a)--(d).
This also suggests that the dissimilarity measure~(\ref{eq:d})
achieves its intended purpose.

The symmetric pairs with high dissimilarity are usually formed by one distribution with strong
peak(s) and another displaying low variability or small peaks.
Figures~\ref{fig:complete}(e)--(f) show the distance distributions
for two symmetric pairs discovered by our procedure.
Especially the distance pattern of $w=CACAGGC$ is noteworthy.
It shows several peaks whose size goes up, which is a very unusual
behavior in distance distributions between words.

\begin{figure}[ht]
\vspace{-0.15cm}
\centering
(a)\includegraphics[width=5.7cm]{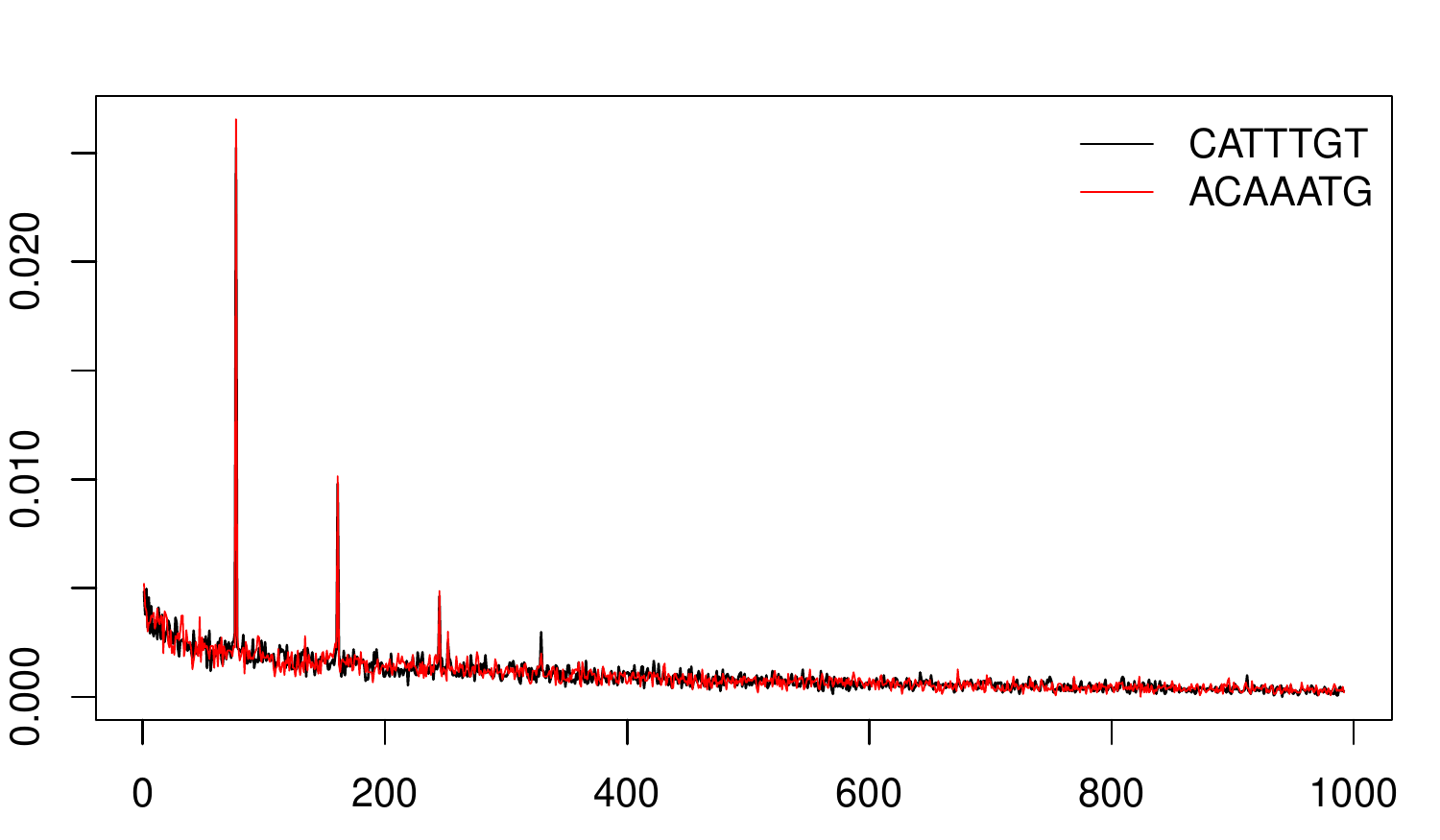}(b)\includegraphics[width=5.7cm]{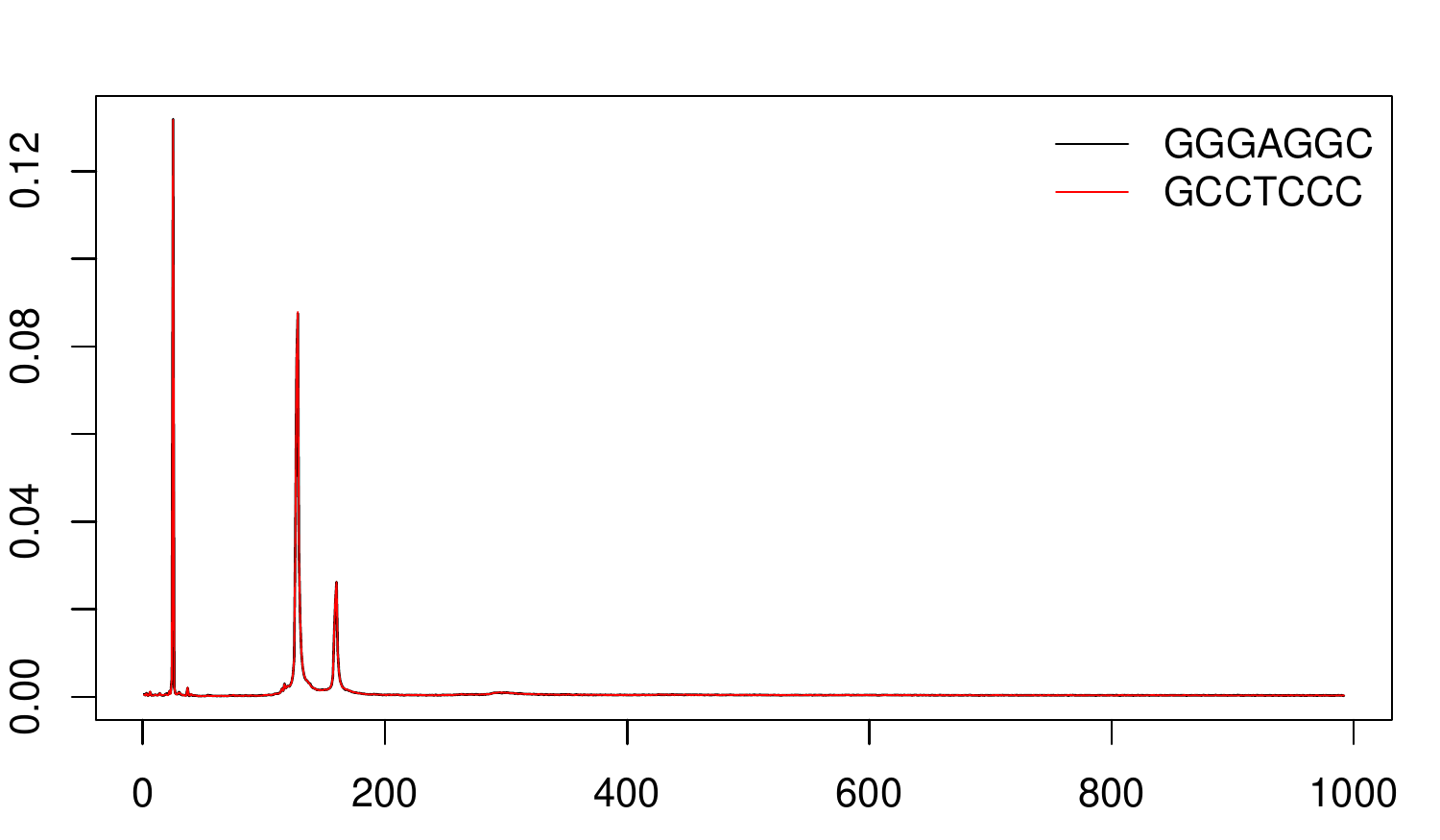}
(c)\includegraphics[width=5.7cm]{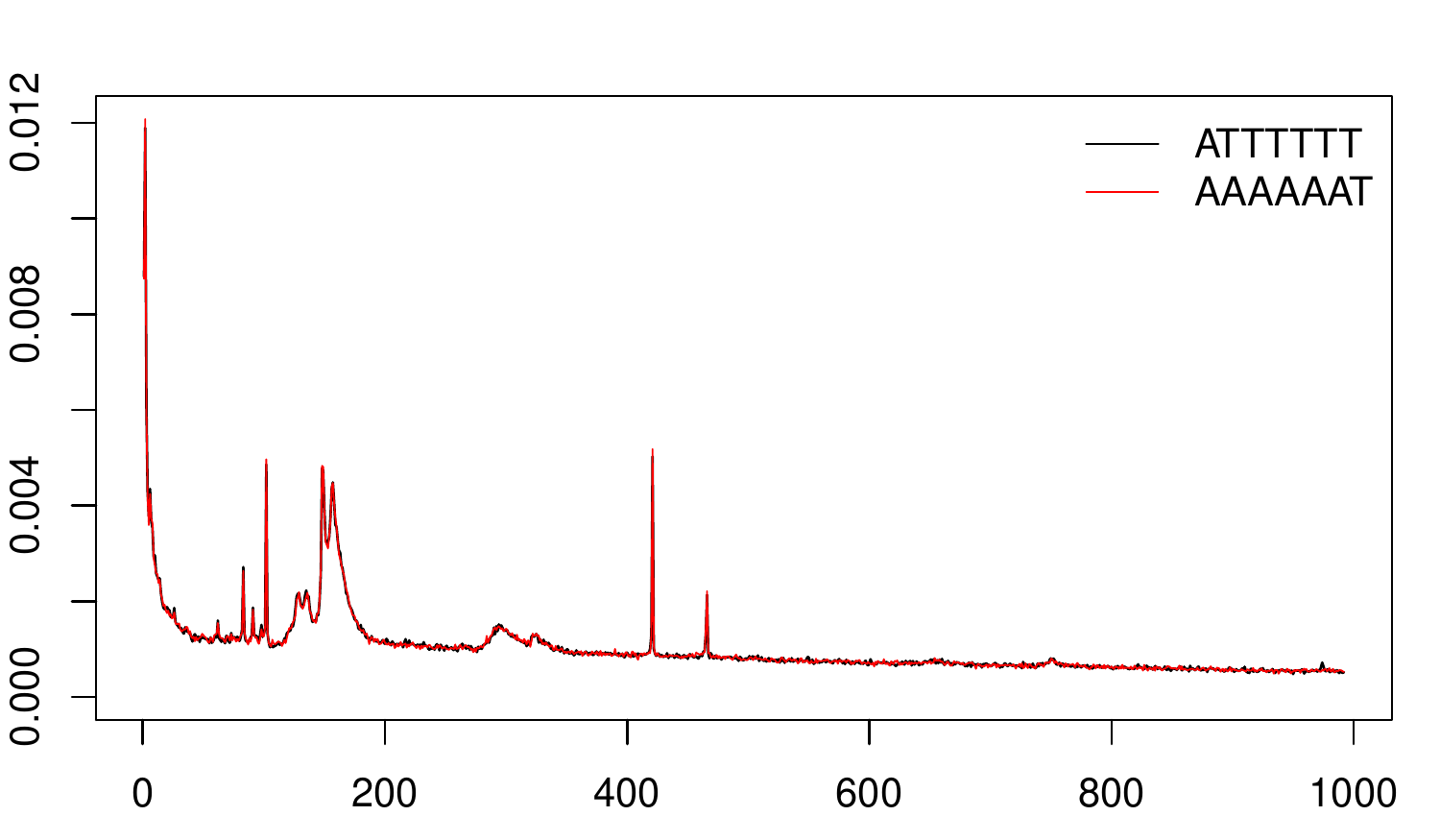}(d)\includegraphics[width=5.7cm]{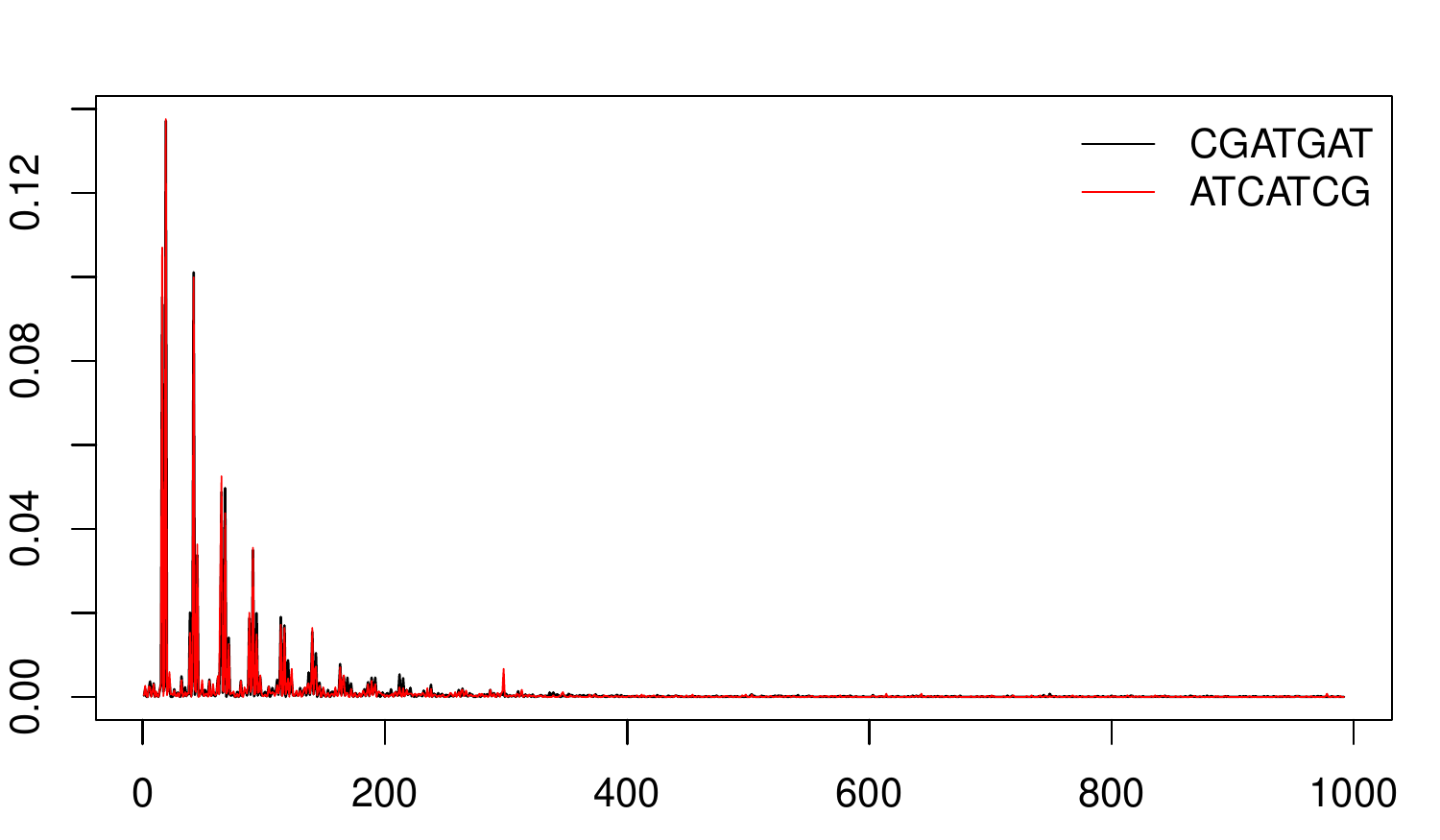}
(e)\includegraphics[width=5.7cm]{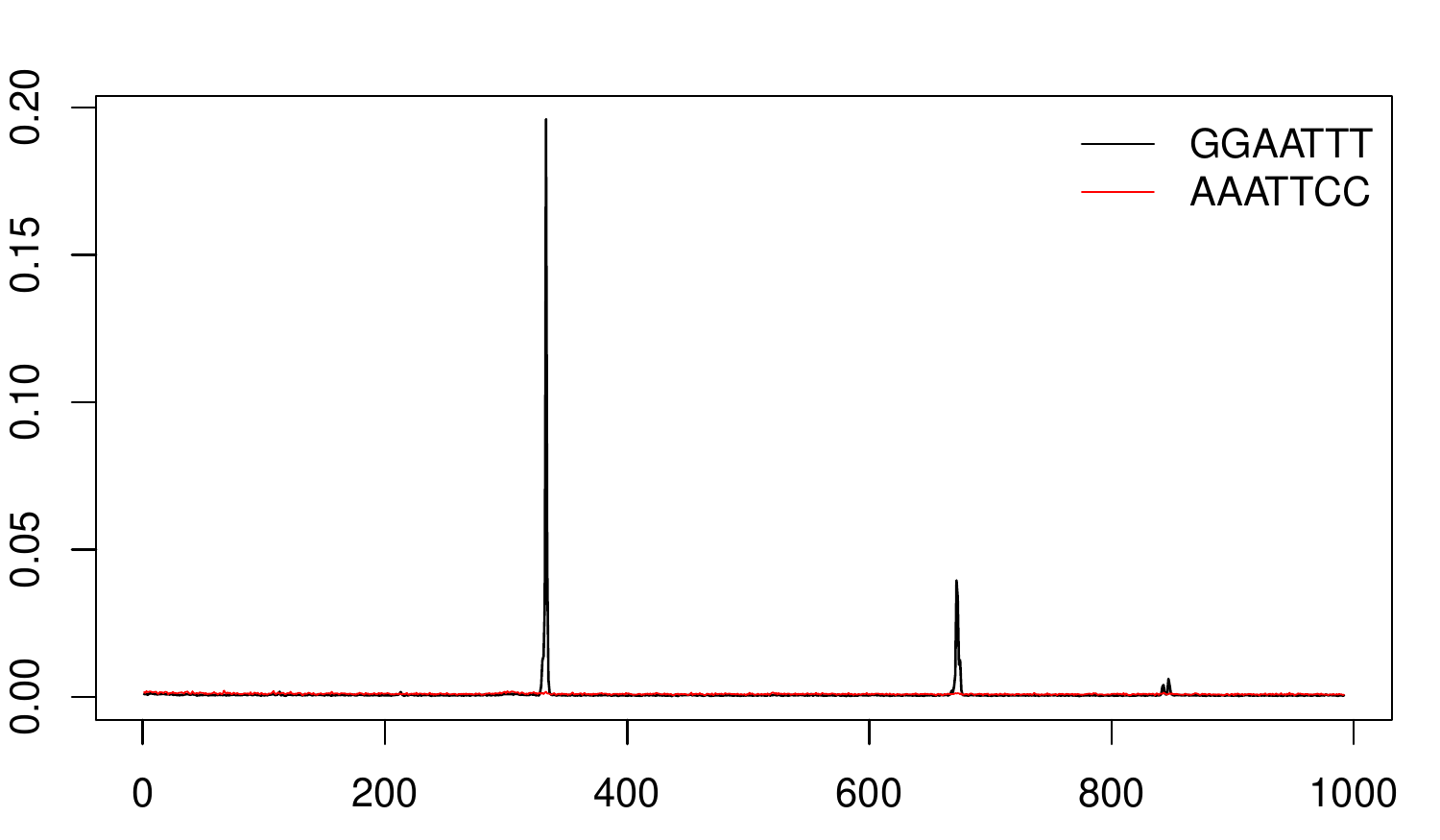}(f)\includegraphics[width=5.7cm]{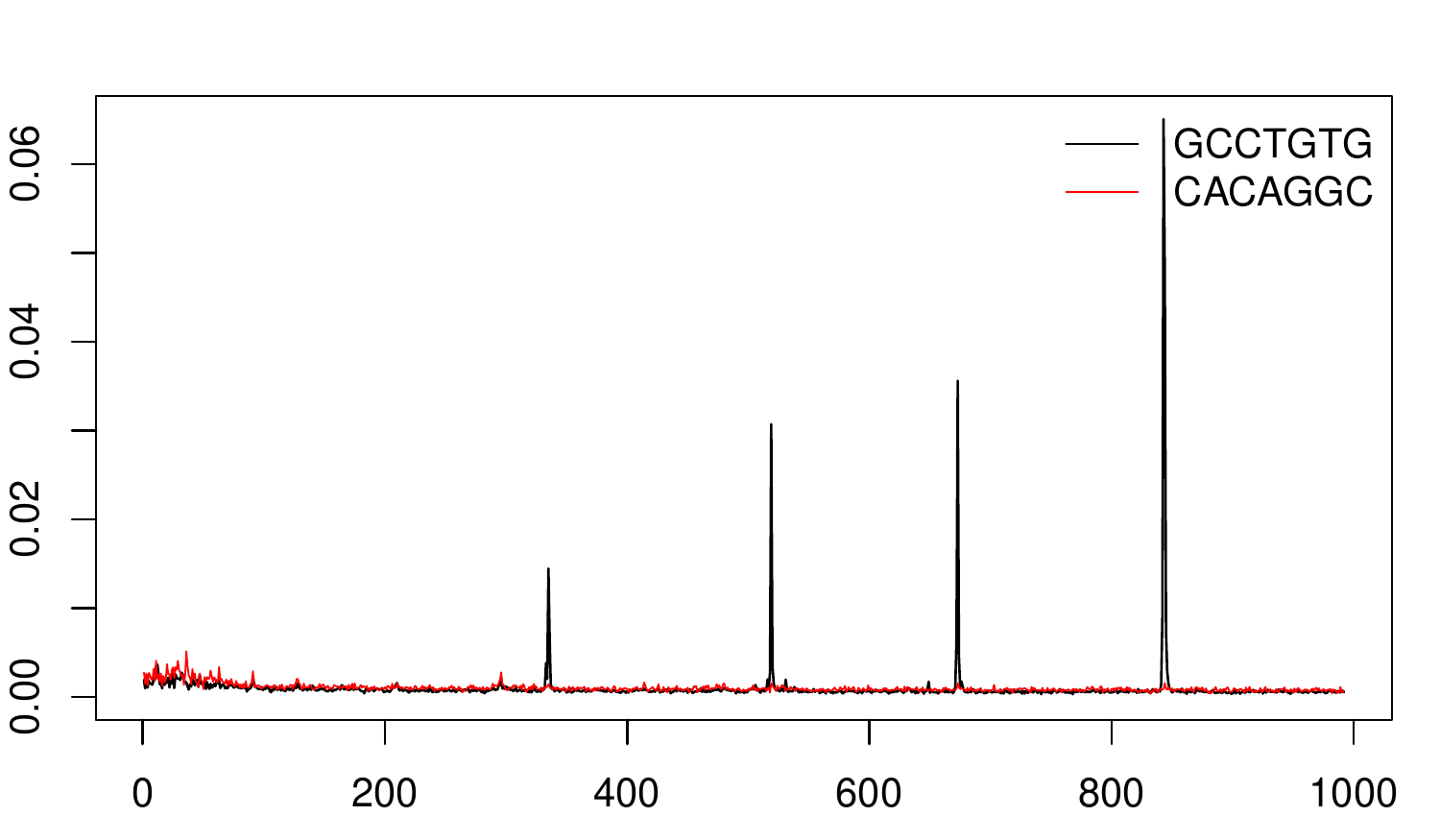}
\caption{Inter-word distance distributions of some reversed complements, $f^w$ and
$f^{\bar{w}}$, with low dissimilarity values: 0.036~(a), 0.003~(b), 0.058~(c),
0.116~(d); and with high dissimilarity values: 178.749~(e), 51.767~(f). Sequence: complete human genome.}
\label{fig:complete}
\vspace{-0.5cm}
\end{figure}

\subsection{Masked Genome Assembly} 

To reduce the effect of repetitive sequences in the original genome
assembly, we also
analyse a masked version of the genome.
All distributions and measures in this subsection
are from the masked sequence.

Masking the genome sequence 
affects the shape of the distance
distributions. Several strong peaks observed in the complete genome are eliminated by masking. For example, the distance distribution
of $w=CACAGGC$
(Figure~\ref{fig:complete}(f)) loses the four strong peaks in the
masked sequence (not shown).

We repeat the previous procedure in the masked sequence, to detect
symmetric pairs whose distance distributions
have very similar or very dissimilar patterns.
The first quartile of $S=\{S^w: w\in \mathcal{A}^k\}$ becomes 546,
so we exclude
the pairs for which $\min\{S^w, S^{\bar{w}}\}\leq 546$,
leaving $D$ with 6075 pairs (see Table~\ref{tab:stats}).

As before we automatically select the symmetric pairs with
dissimilarity below the $10^{th}$ percentile of $D$ (0.328), and those with
dissimilarity above the $90^{th}$ percentile of $D$ (2.454).
The pairs with lowest dissimilarity may be divided in two groups:
those for which both distributions have strong peaks at short
distances, and those whose distributions look like exponential curves
without strong peaks.
These patterns are illustrated in Figure~\ref{fig:mask}(a)--(c).
Interestingly, the unusual pattern of $w=ATCATCG$ in the complete
sequence (Figure~\ref{fig:complete}(d)) remains in the masked
sequence (see Figure~\ref{fig:mask}(b)).

Symmetric pairs with high dissimilarity usually have one
distribution with one or more strong peaks at short distances
(<200) whereas the other has low variability. Some very
dissimilar pairs are shown in Figure~\ref{fig:mask}(d)--(f).

\begin{figure}[ht]
\vspace{-0.5cm}
\centering
(a)\includegraphics[width=5.7cm]{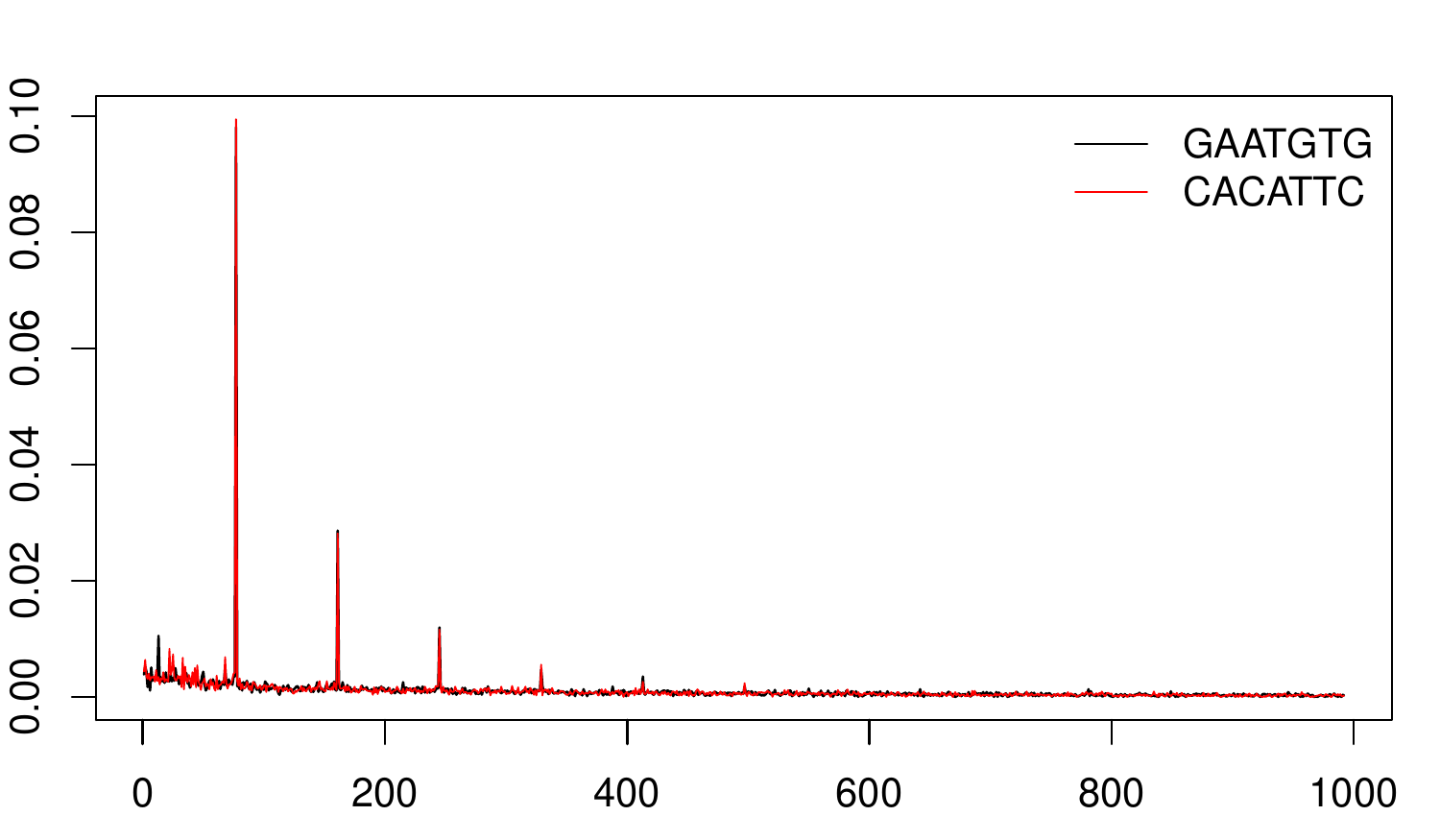}(b)\includegraphics[width=5.7cm]{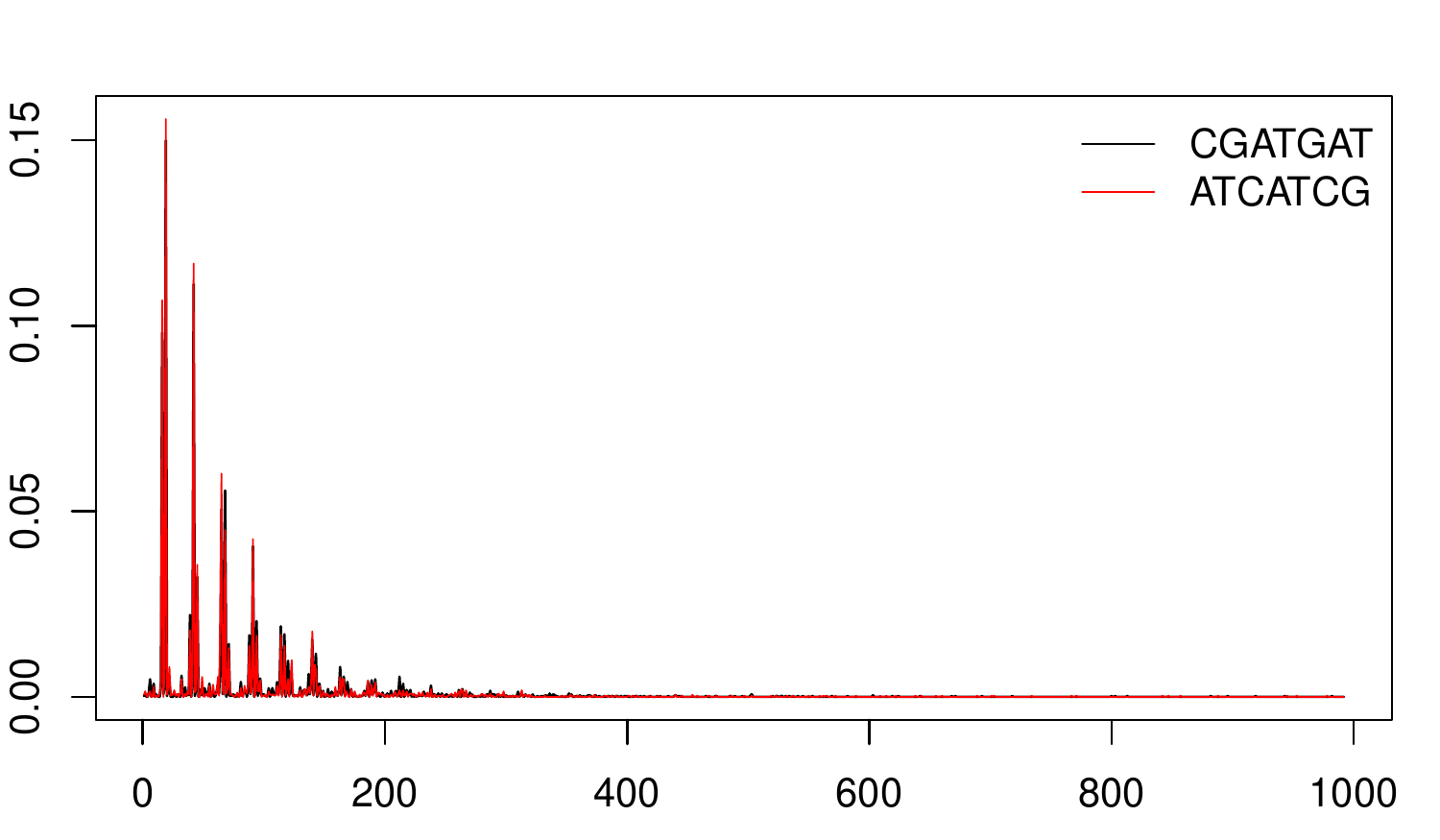}
(c)\includegraphics[width=5.7cm]{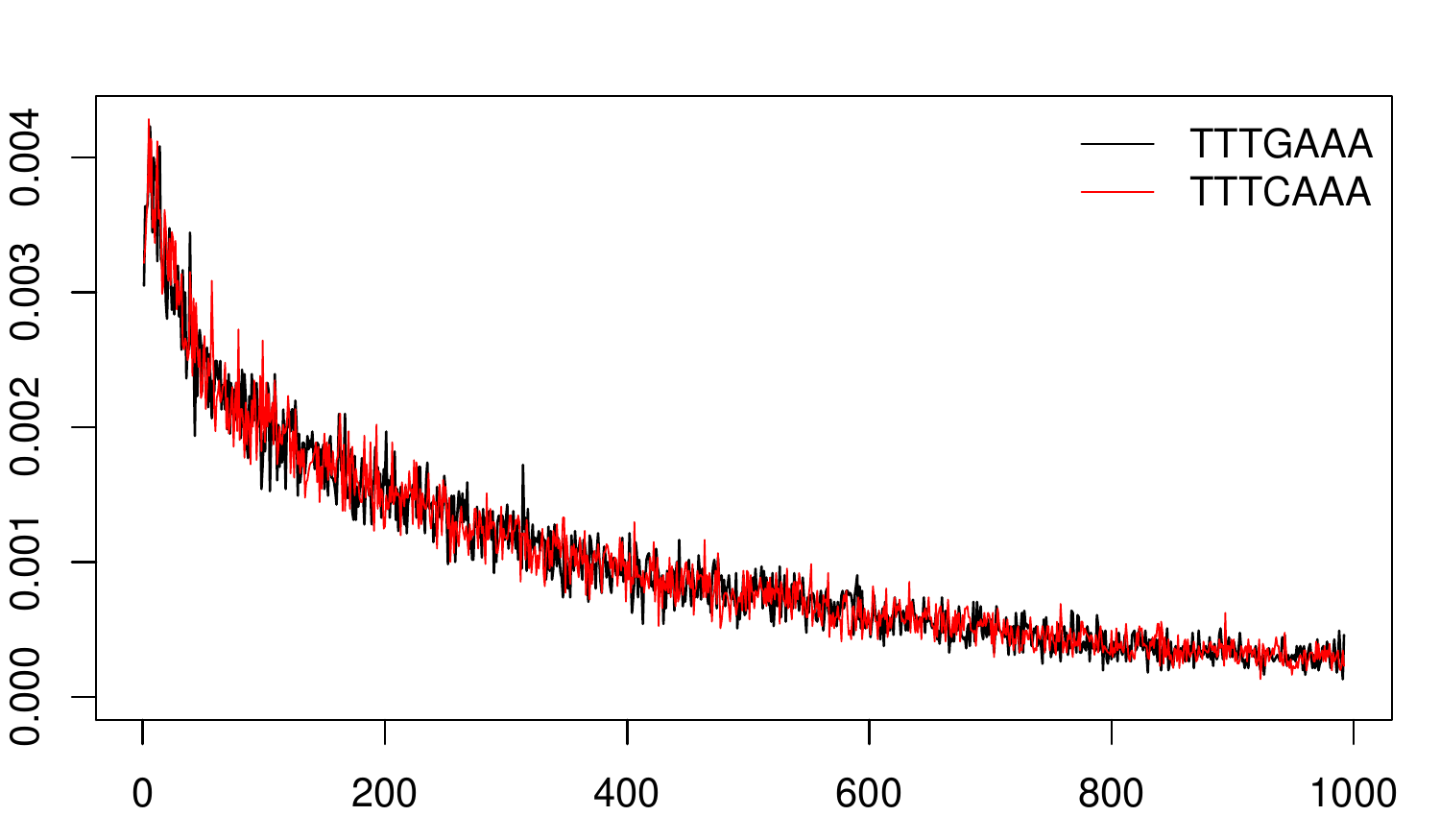}(d)\includegraphics[width=5.7cm]{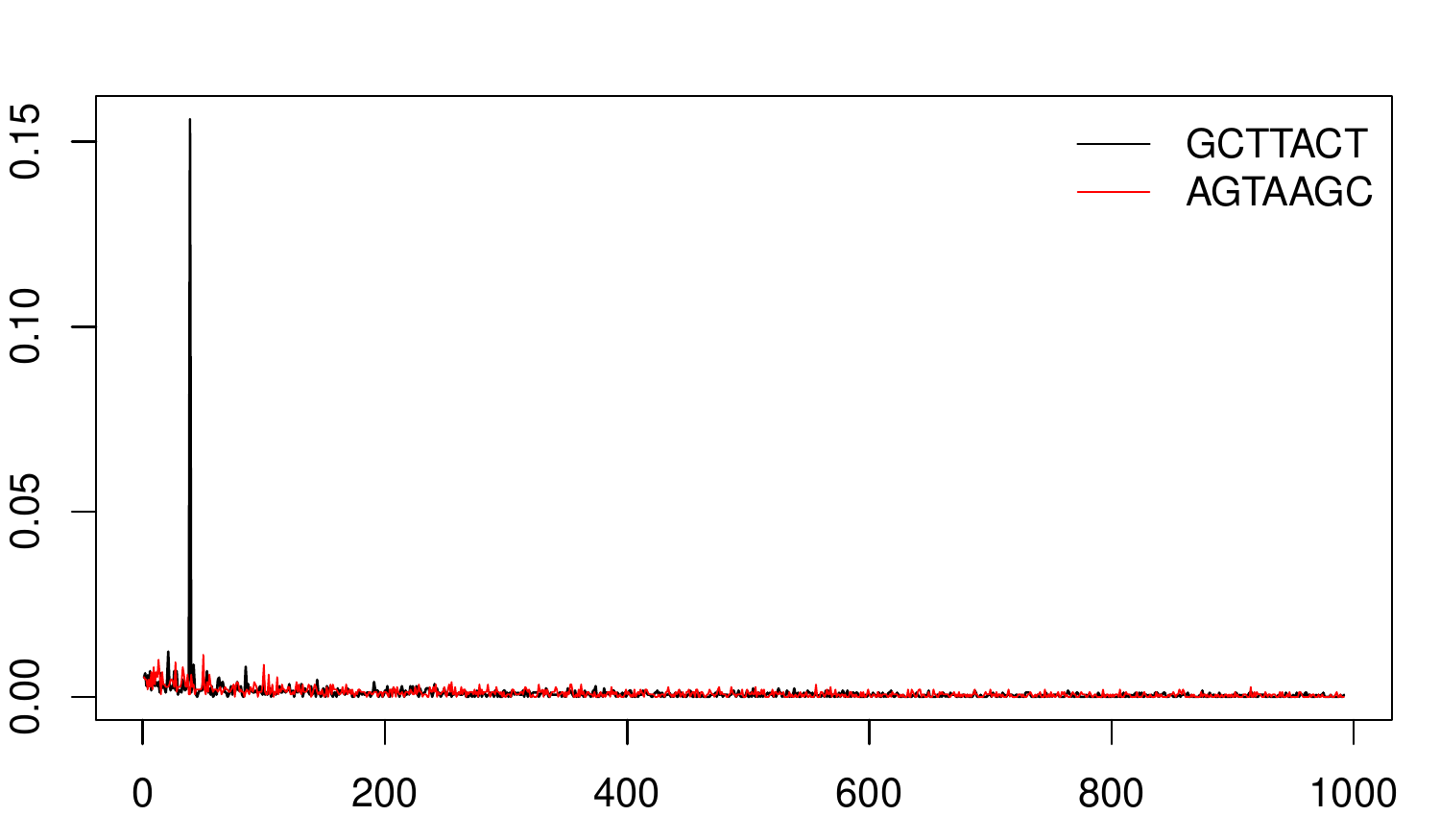}
(e)\includegraphics[width=5.7cm]{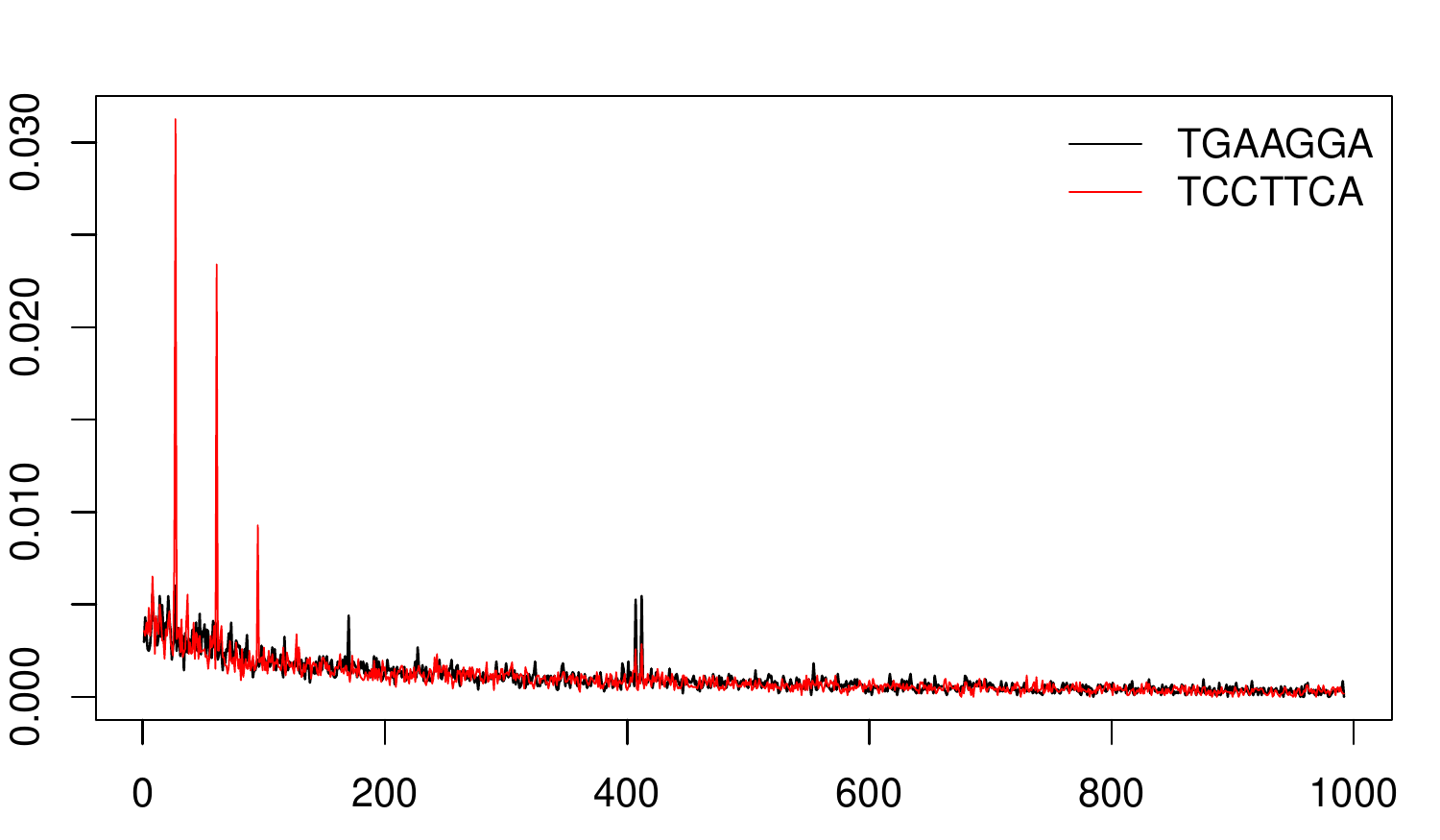}(f)\includegraphics[width=5.7cm]{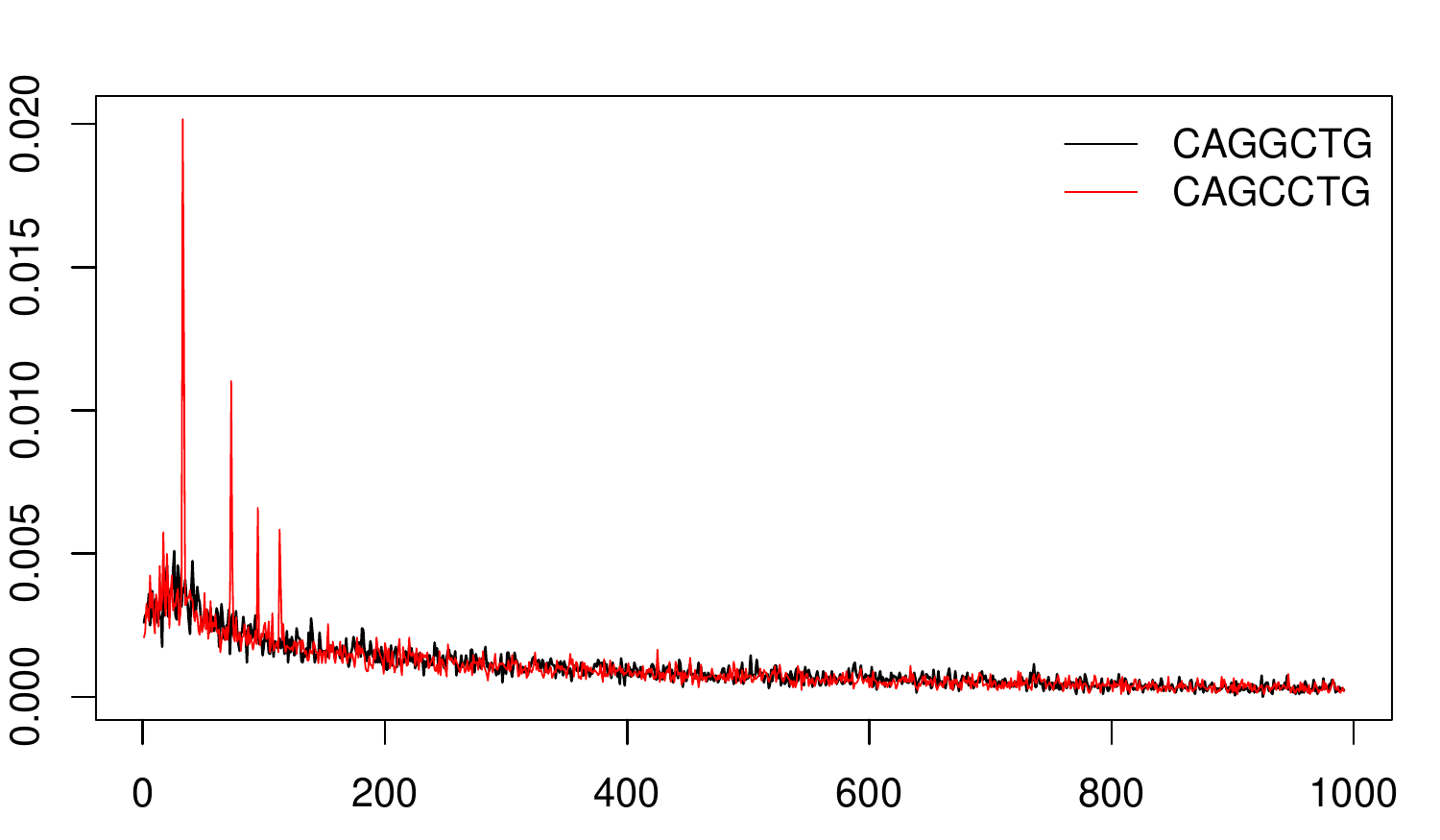}
\caption{Inter-word distance distributions of some reversed complements
with low
dissimilarity values: 0.032~(a), 0.125~(b), 0.144~(c); and with high dissimilarity values: 11.744~(d), 11.310~(e), 6.486~(f).
Sequence: masked human genome.}
\label{fig:mask}
\vspace{-0.5cm}
\end{figure}

To investigate whether an association exists between dissimilar reversed
complements and functional DNA elements, we perform an annotation
analysis for the 15 most dissimilar symmetric pairs.
For each such pair we list the word with the strongest peaks.
Then we look for the `favoured' distance(s), i.e. those where
the strongest peak(s) are located.
These peaks are often concentrated in one chromosome rather than
being spread over the entire genome sequence.
Table~\ref{tab:bio} lists the chromosome in which the favoured
distances are most pronounced, for each of the 15 pairs.
The positions of the words occurring at that distance from
each other are recorded.
Then, we retrieve annotations within these genomic coordinates from UCSC
GENCODE v24.
Interestingly, the words we obtained that are located on chromosome 13
all fall within the
gene LINC01043 (long intergenic non-protein coding RNA 1043)
and all of our words on chromosome 1 are contained in the
gene TTC34 (tetratricopeptide repeat domain 34).
These results suggest that the most dissimilar distributions may be
related to repetitive regions associated with RNA or protein structure.

A deeper investigation into the biological meaning of these words is necessary to investigate whether the observed dissimilarities reflect
the selective evolutionary process of the DNA sequence.

\begin{table}[ht]
\vspace{-0.15cm} \centering \caption{The 15 most dissimilar symmetric
pairs with $k=7$, characterized by their word with the strongest peaks.
The chromosome on which these peaks are prominent is listed.
Masked sequence.}
\setlength\tabcolsep{2.5pt}
\scriptsize
    \begin{tabular}{|l|cc|c|c|c|c|}
		\toprule
    chromosome & \multicolumn{2}{|c|}{13}  & 1      & 4      & 3      & 8 \\
    \midrule
    word $w$   & $ACCATTC\:$ & $GGTAAGC\:$ & $AGCATCT\:$ & $GTTGGTA\:$ & $TGGTATG\:$ & $GCTTACT\:$ \\
           & $CTTCAGG\:$ & $TAAGCAT\:$ & $GAGCATC\:$ & $TGGTAGA\:$ & & \\
           & $GACCATT\:$ & $TCAGGAT\:$ & $TGAGCAT\:$ &        &  & \\
					 & $TCCTTCA\:$ & $TTCAGGA\:$ & & & & \\
    \bottomrule
    \end{tabular}%
  \label{tab:bio}%
  \vspace{-0.15cm}
\end{table}%

\section{Conclusions}

We propose a new dissimilarity measure between distance distributions,
based on discrepancies between their peaks.
Here we use it to evaluate the dissimilarity between reversed complements.

In the complete human genome, we confirm the expected existence of
many symmetric pairs with low dissimilarity, both in word frequency
and in distance distribution.
Even an irregular distribution with strong peaks
is often very similar to that of its reversed complement.
However, our main interest lies in using the proposed dissimilarity
measure to detect symmetric pairs with highly distinct distributions.
In such cases, one of the distance distributions typically has well
defined peaks and the other has low variability.

We also investigate how well our results hold up in the masked
sequence, which excludes major known classes of repeats.
Even though masking generally reduces the dissimilarity between
distributions of symmetric pairs, there remain quite a few word pairs
with high dissimilarity, which in our study was mainly localized on a
specific chromosome and even a specific gen.
A question worth investigating is to what extent the high
dissimilarities may be linked to evolutionary processes.


\section*{Acknowledgment}

This work was partially supported by the Portuguese Foundation for Science and
Technology (FCT), Center for Research \& Development in Mathematics and Applications (CIDMA),
Institute of Biomedicine (iBiMED) and Institute of Electronics and Telematics
Engineering of Aveiro (IEETA), within projects: UID/MAT/04106/2013, UID/BIM/04501/2013
and PEst-OE /EEI/UI0127/2014.
A. Tavares acknowledges the PhD grant from the FCT: PD/BD/105729/2014.
The research of P. Brito is financed by the ERDF - European
Regional Development Fund through the Operational Programme for Competitiveness and
Internationalisation - COMPETE 2020 Programme within project POCI-01-0145-FEDER-006961,
and by the FCT as part of project UID/EEA/50014/2013. The research of J. Raymaekers and
P. Rousseeuw was supported by projects of Internal Funds KU Leuven.



\end{document}